\documentclass[preprint,authoryear,11pt,3p]{elsarticle}

\usepackage{amssymb}
\usepackage{tikz}
\usepackage{graphicx}
\usepackage{caption}
\usepackage{mathtools}
\usepackage{amsmath}
\usepackage{multirow}
\usepackage{colortbl}
\usepackage{color}
\usepackage{tabulary}
\usepackage{bm}
\usepackage{eqparbox}
\usepackage{tabularx}
\usepackage[utf8]{inputenc}
\usepackage{booktabs}
\usepackage{kpfonts}
\usepackage{rotating}
\usepackage{enumitem}
\usepackage{amsthm}
 
\journal{Journal of Operations Management}

\begin{document}

\begin{frontmatter}

\title{Hierarchies Everywhere -- Managing \& Measuring Uncertainty in Hierarchical Time Series}

\author[label3,label1]{Ross Hollyman}
\ead{r.a.hollyman@exeter.ac.uk}
\author[label1]{Fotios Petropoulos\corref{cor1}}
\cortext[cor1]{Corresponding author: f.petropoulos@bath.ac.uk; fotios@bath.edu}
\ead{f.petropoulos@bath.ac.uk}
\author[label2]{Michael E. Tipping}
\ead{mt821@bath.ac.uk}
\address[label3]{Exeter Business School, University of Exeter, United Kingdom}
\address[label1]{School of Management, University of Bath, United Kingdom}
\address[label2]{Department of Mathematical Sciences, University of Bath, United Kingdom}
\date{September 2022}

\begin{abstract}
We examine the problem of making reconciled forecasts of large collections of related time series through a behavioural / Bayesian lens. Our approach explicitly acknowledges and exploits the `connectedness' of the series in terms of time-series characteristics and forecast accuracy as well as hierarchical structure. By making maximal use of the available information, and by significantly reducing the dimensionality of the hierarchical forecasting problem, we show how to improve the accuracy of the reconciled forecasts. In contrast to existing approaches, our structure allows the analysis and assessment of the forecast value added at each hierarchical level. Our reconciled forecasts are inherently probabilistic, whether probabilistic base forecasts are used or not.  
\end{abstract}
\end{frontmatter}

\section{Introduction}

\emph{``...we have suggested that people are strongly biased in favour of the inside view, and that they will normally treat significant decision problems as unique even when information that could support an outside view is available''} \citep{Kahneman1993-xu}

In their 1993 article Kahneman and Lovallo focus on cognitive biases affecting forecasting and decision making problems in organisational contexts where both problem specific (`inside') and problem class (`outside') information are available to the forecaster. The inside view treats each forecasting problem as unique, and focuses on the specifics of each case in isolation, whereas the outside view focuses instead on a broader set of statistics derived from cases judged to be similar in key respects to the problem at hand.

\cite{Kahneman1993-xu} explicitly limit the focus of their article to forecasting and decision making problems which are largely one-off in nature and where appropriate analogues for outside modes of forecasting are difficult to tie down. They expect that more routine, repeatable forecasting exercises are able to be managed in such a way that eliminates or substantially reduces judgemental bias. In this article, however, we examine the problem of forecasting ensembles of time-series in exactly this (routine, repeatable) scenario, and in particular on scenarios where aggregates of various subsets of the ensemble are also of managerial interest. Such Hierarchical Forecasting (HF) scenarios have received much attention in the (mostly quantitative) forecasting literature in recent years \citep[see, for example,][and references therein]{Athanasopoulos2009,Hyndman2011,Wickramasuriya2019,Panagiotelis2020,HollymanRoss2021Ufr,Di_Fonzo2022-be,PANAGIOTELIS2022}, and are also the subject of a study by \cite{Kremer2015} which considered experimental evidence on the behavioural biases of judgemental forecasters in similar scenarios.

We suggest that most current academic quantitative forecasting practice embeds exactly the same bias towards the inside mode of forecasting identified by \cite{Kahneman1993-xu}, even though each such ensemble is by definition a ready made collection of analogues which enables an outside view to be readily developed.

This paper focuses primarily on measuring and managing uncertainty in the context of forecasting multivariate time series which are arranged in a hierarchical structure. Although several recent papers address this topic, the approach we describe below has more in common with the approach set out in \cite{West1997}, which predates by several years the recent interest in this topic.

Our approach focuses in particular on three different types of uncertainty which are largely brushed over by the recent literature. Firstly, we produce forecasts which are inherently probabilistic. Whilst several papers \citep{Jeon2019,BenTaieb2020,PANAGIOTELIS2022} have to some extent addressed this topic, their focus is on reconciling probabilistic base forecasts. Our focus is instead \emph{probabilistic forecast reconciliation} - our base forecasts may or may not be probabilistic, but reconciled forecasts are always so. Our approach generates probabilistic \emph{reconciled} forecasts whether base forecasts are probabilistic or not.

Secondly, we focus on the the role of correlations between series in determining how shocks affecting different parts of the hierarchy are propagated through the entire system. Univariate forecasting models are (rightly) very popular and have an excellent track record in practice when univariate forecast accuracy is the sole assessment criteria. Whilst the benefits of univariate accuracy carry over to multivariate settings, additional factors often come in to play when risk / uncertainty need to be taken in to account. Whilst the expected value of an aggregate is simply the sum of the values of the constituents, the same is not true for the variance.  From a management perspective, it is often important to quantify risk across a portfolio of time-series or, more crudely put, to take in to account the extent to which all forecasts are wrong at the same time. To illustrate this point, we adopt the example of \cite{West1997}, and references therein, who examine a scenario where a collection of forecasts are impacted by a common factor. Assume demand for a collection of $i=1...1000$ products is described as:
\begin{equation}y_i = x_i + f + e_i\end{equation}

\noindent where $x_i$ is the (assumed known) mean demand for product $i$, $f \sim N(0,1)$ and $e_i \sim N(0,99)$. The $e_i$ are assumed uncorrelated, so that $Cov(x_i,x_j)=0$ for all $i$ and $j$. The variance of $y_i$ is then simply 1+99 = 100. Precise knowledge of the value of $f$ makes very little difference to predictions of $y_i$, reducing variance by only 1\%, i.e, $Var(y_i|f)=99$. Set the total series 
\begin{equation}
T = \sum^{1000}_{i=1} y_i
\end{equation}

\noindent so that $E(T) = \sum E(X)$ but $Var(T) = Var(1000 \times Var(f) + 1000 \times Var(e_i)) = 10^6 + 99,000$ so that for series T the variance of $f$ swamps the contribution of the individual series. 
This is the same effect as diversifying stock specific risk in a portfolio of equity securities whilst leaving market risk broadly untouched on which the Capital Asset Pricing Model is founded. At the top level, knowledge of $f$ makes a significant difference to the variance of $T$, with $Var(T|f)=99,000$.
Ignoring cross-series correlations can therefore have significant impacts on risk when decisions are aggregated. The same basic result applies when more than one factor affects the set of series, as is highly likely in a large hierarchical setting. In this context, correlations between the factors also come in to the equation. Considerations of this nature lie behind the almost ubiquitous reliance on multi factor risk models in portfolio management; we deploy the same techniques in our setting.

Thirdly, we note that HF models are extremely `parameter heavy' - both in terms of the statistical estimation underpinning the reconciliation process \citep{HollymanRoss2021Ufr,Pritularga2021-gl}, but also in the parameters underpinning generation of the base forecasts themselves -- a point largely ignored in the existing literature. HF is fundamentally a forecast combination process which inherently mitigates estimation error \citep{Elliott2013} and several flavours of HF shrink the high dimensional covariance matrix of forecast errors in different ways. The most popular approaches are the use of ordinary least squares (OLS) or weighted least squares (WLS) - shrinking off-diagonal covariance terms to zero, or use of the general covariance matrix shrinkage algorithms \citep[e.g.,][]{Wickramasuriya2019}. The approach we describe in this paper directly confronts this issue, and in particular we use information embedded in the hierarchy, tying risk explicitly to common factors implied by the hierarchy.

The approach we adopt has several other advantages over recently developed methodologies. Whilst significantly more parsimonious in terms of parameter estimation, the parameters that we do estimate have a clear interpretation and provide an analytical framework that aids monitoring and refining the forecasting process. We can, for example, determine the value added by forecasts at different hierarchical levels, allowing us to focus resources on forecasting more accurately, or perhaps choosing not to generate base forecasts at all in some cases. Existing approaches focus on the forecast error covariance matrix and require exactly one forecast for each time series. Our approach is more general, and allows several models for each level, and for no base forecasts to be made at all at some levels. For this reason, the approach readily extends to temporal and cross-temporal hierarchies. Our solution also copes with missing data and unbalanced panels with little difficulty.

In short, our solution utilises hierarchies everywhere - we specify a prior over the entire collection of series, defined by a multivariate model for the atomic series, which are then aggregated via the usual HF summing matrix. This process defines a prior mean and variance for the entire collection, the series covariance terms being specified in terms of a hierarchical factor model. Using a hierarchical prior structure, we then calibrate the set of base forecasts and update the prior. To achieve computational efficiency, and to enable interpretation and understanding of the model, we prepare one set of reconciled forecasts per hierarchical level and then combine these in the final stage of the modelling process. The algorithm is wrapped up in an efficient Markov Chain Monte Carlo (MCMC) sampling process which averages over likely values of the set of model parameters. In the next section, we describe the various components of the model, before moving on to describe the computational details of the sampling algorithm in section 3. In section 4, we illustrate the application of our model using a well known data-set. Section 5 concludes.

\section{The model}

In this section, we describe the various components of the model, and explain how they fit together to produce a novel solution to the HF problem. Computational details are left to the following section.

\subsection{Notation}

In what follows, we employ the following notation. We consider a collection of $m$ time series, observed over a time interval of length $T$. Of the $m$ series, $n$ are defined to be bottom level or 'atomic' series, and $m^* = m-n$ are aggregates of various subsets of the $n$ atomic series defined by the summing matrix of the hierarchy. We prefer the term `atomic' for the what were initially referred to in the literature as bottom level series for the reasons set out in \cite{Panagiotelis2020}: HF techniques are applicable not just to hierarchies, but in any setting where linear combinations of subsets of series are of incremental interest. We denote the observations of the atomic series as $b_{it}$ $i=1...n$ so the $m$ vector of observations at time $t$ is $\mathbf{y}_t = \mathbf{S}\mathbf{b}_t$ where $\mathbf{S}$ is the usual HF summing matrix. Again, as is customary in the literature, we denote the `base forecasts' of $\mathbf{y_t}$ as $\mathbf{\hat{y}}_t$. Additionally we write the prior expectation of $\mathbf{b}_t$ (after estimating the multivariate time series model, but before updating with the forecasts) as $\mathbf{\underline{b}}_t$ so that $\mathbf{\underline{y}}_t = \mathbf{S\underline{b}}_t$. In order to accord with the rest of our notation, we write the posterior mean of $\mathbf{y}_t$ as $\mathbf{\bar{y}}_t$ (as opposed to $\mathbf{\tilde{y}}_t$ which is conventional in the HF literature). We denote the (scalar) observation residuals (relative to the prior) as $r_{it} = y_{it} - \underline{y}_{it}$, and the base forecast residuals (again defined relative to the prior) as $g_{it} = \hat{y}_{it} - \underline{y}_{it}$. The terms $\mathbf{r}_t$ and $\mathbf{g}_t$ are then vectors of these quantities at time $t$. We will also sometimes refer to a $T$ vector of observations of a given series $i$ as $\mathbf{r}_i$

\subsection{Exchangeable time series}

A core idea is that time series arranged in a hierarchy will very often share common time series characteristics, especially so once differences in location and scale are accounted for. Most statistical time series models have at their core an assumption that today and tomorrow's observations are similar to or `exchangeable' \citep{GelmanAndrew2013Bda,Goldstein2007} with those of the past. We argue that in many hierarchical situations it is sensible \emph{a priori} to make exactly the same assumption in the cross-section. That is to say that there exist common patterns in the time series evolution of the various series, and we can potentially make gains in predictive accuracy by taking this in to account during the modelling process. Approaches based on this idea have recently become popular in the forecasting and machine learning literature where they are often referred to as `cross-learning' and are central to Bayesian Hierarchical modelling approaches which have been in use for many years. The Bayesian approach embodies Occam's razor - not multiplying entities (parameters) unnecessarily. `Local' parameters only differ from 'global' values where justified by the data. \cite{Kahneman1993-xu} suggest that because deliberately disregarding `inside' information can appear irresponsible, ``... a deliberate effort will therefore be required to foster the optimal use of outside and inside views in forecasting''. The Bayesian approach we adopt here seeks to optimally blend \emph{both} approaches.

This concept applies to hierarchical series in two important ways. Firstly, and as described above, the parameters of the base time-series forecasts, especially where these are generated using the same class of models, might be expected to have some degree of commonality. For example where exponential smoothing models are fitted using some kind of automated algorithm, differences in fitted smoothing parameters are likely to represent both variation in true values across series, plus an element of estimation error. The estimation error can (and should) be reduced during estimation by shrinking the parameters towards global average values, although this is often not done in practice. We address this issue by proposing a prior expectation for the future values of all of the atomic series in the hierarchy based on the exchangeable time series models described in \cite{West1997}, \cite{PradoRaquel2021} and references therein. These state-space time-series models propose that a common state is shared across all series in a collection. Both forecast errors and state evolution variances are effected by the series covariance matrix, so correlated series will exhibit commonalities in both expected (state evolution) and unexpected (observation shocks) behaviour. If such a model is defined for the atomic series in a hierarchy, then values for the entire collection can be obtained simply via the HF $\mathbf{S}$ matrix. We then update our expectations from the purely exchangeable model with a set of base forecasts from some external source.

Secondly, we consider commonality in the predictive accuracy of the models underlying these base forecasts. Where forecasts are produced for similar time-series using similar algorithms, one might expect forecast accuracy (however measured) to cluster around some central value. Again, departures from this value will be driven in part by fundamental characteristics of the series in question and the particular model form, but partly by estimation error. This observation helps us to optimally revise our predictions of future outcomes in the light of observing the base forecasts. We achieve this via a forecast calibration process, to which we now turn.

\subsection{Forecast calibration}

The calibration process aims to estimate the incremental value added of the base forecasts over the purely exchangeable prior. We follow most of the HF literature where in-sample fitted values are used to estimate forecast combination weights which are then applied to out-of-sample forecasts, but reflect this in our prior by shrinking the incremental forecasting ability of the more complex (in terms of number of parameters) models towards zero. We also reflect the expectation that the forecasting power of models generated from the same source (be this a forecasting algorithm or a judgemental process) will tend to cluster around some common value. The degree of shrinkage in practice is driven by the data, and using a Bayesian hierarchical approach we are able to neatly incorporate both the expected patterns in cross sectional predictability and a preference for simpler over more complex models.

\subsection{Linear Bayes Estimation}

Our workhorse approach is Linear Bayes Estimation (LBE) described in \cite{Goldstein2007} and the references therein, and used as the cornerstone of the forecast refinement methodology for active portfolio management in \cite{GrinoldKahn}. The same basic approach emerges from the literature on expert opinion analysis and `Bayesian Predictive Synthesis' (BPS) \citep{Lindley1979-ij,Genest1985-qb,West1988-rp,West1992-ml,West1992-ty,McAlinn2016-pc,Johnson2018-bh}. LBE constitutes a very general Bayesian learning approach, and is useful in an array of situations where specifying a full prior distribution for every part of the model is impractical.

The approach works as follows. Given an $n$ vector of quantities $\mathbf{r}$, and a set of forecasts of $\mathbf{r}$ contained in an $m$ vector $\mathbf{g}$, the revised expectation of $\mathbf{r}$ given $\mathbf{g}$ is:

\begin{equation}
    \label{LBE-Mean}
    E(\mathbf{r}|\mathbf{g}) = E(\mathbf{r}) + Cov(\mathbf{r},\mathbf{g})Var^{-1}(\mathbf{g})(\mathbf{g}-E(\mathbf{g}))
\end{equation}

The $n \times n$ variance/covariance matrix of $\mathbf{r}$ given $\mathbf{g}$ is:

\begin{equation}
    \label{LBE-Var}
    Var(\mathbf{r}|\mathbf{g}) = Var(\mathbf{r}) + Cov(\mathbf{r},\mathbf{g})Var^{-1}(\mathbf{g})Cov(\mathbf{g},\mathbf{r})
\end{equation}

In order to update of $\mathbf{r}$ given $\mathbf{g}$ in empirical settings, we need to estimate $Var(\mathbf{g})$ and $Cov(\mathbf{r,g})$. Whilst this is a relatively simple matter when $n$ and $m$ are small, it can be a daunting task in large hierarchies. We now develop a scalable approach to this problem. 

\subsubsection{A univariate analysis}

Consider $t=1...T$ observations of a \emph{univariate} time series $r$, and $g$, a set of forecasts of $r$ from some external source. We can write the outcomes as the sum of a linear function $f$ of the forecast and an error term:

\begin{equation}
    r_{t} = f(g_{t}) + e_{t}
\end{equation}

The coefficient of determination or $R^2$ associated with this relationship then gives the proportion of the variance of $r$ explained by the forecasts $g$. We assume this is equal to the (squared) correlation coefficient of the forecasts and the outcomes:

\begin{equation}
    \label{r2}
    R^2 = \rho^2 = \frac{Var(g)}{Var(r)}
\end{equation}

To estimate the accuracy of the forecasts in practice we can fit the regression:

\begin{equation}
    \label{ri}
    r' = \rho g' + e
\end{equation}

Where  $r'$ and $g'$ denote standardised data $r$ and forecasts $g$ so $r'$ and $g'$ have mean zero and standard deviation of 1.

In a financial context, this equation is usually applied to returns where the zero mean assumption is generally appropriate, at least in the short term. In a HF context we apply this equation to residuals from a base-line multivariate time series model which we assume applies to all series in the collection, as discussed above. Biases in $g$ can be accounted for by adjusting this regression to allow for a constant, but for clarity we assume in what follows that $g$ is unbiased. Estimation by univariate regression in this way is convenient and opens the door to shrinking the estimate of $\rho$ using the standard Bayesian regression model. Such models are then trivially extended using a hierarchical prior to capture commonality in $\rho$ across sets of similar time series, and we can (and do) easily enforce the additional constraint that $0\geq \rho \geq 1$ as described in \cite{Geweke1996-vv}.  As $\rho$ is scale free, we can write down an equation for $g'$ in terms of $r'$:

\begin{align}
    \label{rho_gen1}
    g' &= \rho r' + \epsilon\\
    \epsilon & \sim N(0,\sigma^2_\epsilon)
\end{align}

So that the original $r$ and $g$ (before standardising) are related by:

\begin{equation}
    \label{rho_gen2}
    g = \rho r \frac{{\sigma_g}}{{\sigma_r}} + {{\sigma_g}} \epsilon
\end{equation}

\noindent with the respective standard deviations of $r$ and $g$ denoted by ${{\sigma_g}}$ and ${{\sigma_r}}$. From (\ref{r2}) we can write ${{\sigma_g}}=\rho{{\sigma_r}}$, and from (\ref{rho_gen1}) we see that $\sigma^2_{\epsilon} = \sqrt{1-\rho^2}$. Substituting these in to \ref{rho_gen2} gives:

\begin{align}
    \label{gen-model}
    g &= \rho^2 r + \rho \sqrt{1-\rho^2} \sigma_{r} z\\
    z &\sim N(0,1)
\end{align}

\noindent where $z$ is random Gaussian noise assumed uncorrelated with $r$. This equation represents a generative model for the forecasts $g$ \citep{Murphy2012-za}. The focus is on the joint distribution of g and r, from which $r|g$ is then estimated, in contrast to the `discriminative' approach taken by the MinT approach \citep{Wickramasuriya2019} which seeks to estimate $E(r|g)$ directly. Equation (\ref{gen-model}) is utilised initially to the best of our knowledge in \cite{GrinoldKahn}. The model assumes that the univariate forecasts $g$ can be written as a function of $r$ plus scaled noise, where the scaling is a function of the variance of $r$ and the forecast accuracy or calibration parameter $\rho$. \cite{GrinoldKahn} refer to $\rho$ as the 'Information Coefficient' of an investment strategy, and build a comprehensive framework for active portfolio management around it. There is, in our view, much to recommend the use of this approach to the measurement of  forecast accuracy, as it is scale free and relatively easy to interpret, and unlike many popular forecast accuracy metrics can be utilised directly to calibrate forecasts in an intuitive way. Interpreting $\rho$ as a correlation coefficient links  directly to the out of sample $R^2$ \citep{Campbell2007-vb}. Indeed, in a recent work, \cite{Chicco2021-ik} suggest that $R^2$ has properties preferable to most competing metrics in the context of assessing forecast accuracy of regression type models.  

\subsubsection{Multivariate development}

Equation (\ref{gen-model}) can now be used to extend the model to a \emph{multivariate} setting, by deriving expressions for the covariances and variances needed for Linear Bayes updating in terms of the individual $\rho_i$ for each variable and the variance/covariance matrix of the observations $r_i$. This is particularly helpful in the HF setting when we have additional forecasts for aggregates of subsets of a given collection of series, but the procedure is a general one, and applies in principle to any multivariate setting. Making the further assumption that $z_i$ are uncorrelated with all $r_j$ it follows directly from (\ref{gen-model}) that: 

\begin{equation}
    Cov(r_j,g_j) = \rho_j^2 Var(r_j)
\end{equation}

\noindent and for any two time series $r_i$ and $r_j$ that:

\begin{equation}
    \label{cross_forecast}
    Cov(r_j,g_i) =  \rho_i^2 Cov(r_j,r_i) 
\end{equation}

These expressions combine to give the numerator term in the LBE equations (\ref{LBE-Mean} and \ref{LBE-Var}). For the denominator, we require the inverse of the variance/covariance matrix of $\mathbf{g}$. We again assume that $z_i$, $r_j$ are uncorrelated for all $i$ and $j$. It then follows from (\ref{gen-model}) that: 

\begin{equation}
    Cov(g_i,g_j) =  \rho_i^2 Cov(r_i,r_j) \rho_j^2 + \rho_i \sqrt{1-\rho_i^2} \sigma_{r_i} Corr(z_i,z_j) \rho_j \sqrt{1-\rho_j^2} \sigma_{r_j}  
\end{equation}

\noindent or (again given \ref{r2}):

\begin{equation}
    Corr(g_i,g_j) =  \frac{\rho_i}{\sigma_{r_i}} Cov(r_i,r_j) \frac{\rho_j}{\sigma_{r_j}} + \sqrt{1-\rho_i^2} Corr(z_i,z_j) \sqrt{1-\rho_j^2}  
\end{equation}

If the residual forecast errors can be assumed to be independent conditional on the covariance of $r$, i.e. $z_i$ are assumed IID Gaussian, the above simplifies to:

\begin{equation}
    \label{corr_g}
    Corr(g_i,g_j) =  \frac{\rho_i}{\sigma_{r_i}} Cov(r_i,r_j) \frac{\rho_j}{\sigma_{r_j}} = \rho_i Corr(r_i,r_j) \rho_j 
\end{equation}

\noindent i.e., the correlation structure of the forecasts, which forms the denominator in the LBE update of the data, can be written as a product of the univariate calibration coefficients $\rho$ and the data correlation matrix. This is a particularly parsimonious form which avoids the need to compute the full $(2m)^2$ covariance matrix of forecasts and outcomes. We now turn to a factor model representation of the variance of $r$ which further reduces the dimensionality of the updating equations.

\subsection{Factor model}

The approach outlined above gives expressions for updating time series arranged in a hierarchy as a function of univariate calibration parameters for each series $\rho$, and estimates of the covariance of the underlying series. For large hierarchies this remains a daunting task. High dimensional covariance matrices are notoriously difficult to estimate, and optimisation/ regression analysis based on them is highly suspect unless proper shrinkage techniques are used. A common approach to this problem in a variety of settings is to use a factor model. Factor models are essentially a data compression/dimension reduction technique which allow large variance/covariance matrices to be estimated using a relatively small number of parameters, and are widely used in econometric and financial setting. Indeed factor models tend to be the standard approach for capturing covariance between securities in portfolio management applications. In a HF setting, defining an exchangeable time series model to account for predictable time series variation in the location of a time series allows the cross-sectional co-variation to be modelled using a similar factor based approach.

Given a vector of zero mean time series $\mathbf{r_t}$ observed at times $t=1...T$, when the variance / covariance matrix of $r$ follows a factor structure, we can write:

\begin{equation}\label{var_g}
    Var(\mathbf{r}) = \boldsymbol{\Delta} \boldsymbol{\Sigma_{r}}\boldsymbol{\Delta'} + \mathbf{D_r} 
\end{equation}

\noindent where $\boldsymbol{\Delta}$ is the $n x q_g$ matrix of factor loadings and $\boldsymbol{\Sigma_{r}}$ is the variance/covariance matrix of the factors driving $\mathbf{r}$. $D_r$ is the $n \times n$ diagonal matrix of the specific variances of $r$. If we assume the factor loadings are constant through time, then conditional on the factors f each series $i$, $i=1...n$ can be written as a regression equation:

\begin{equation}\label{factor-fit}
    r_{it} = \boldsymbol{\Delta} \boldsymbol{f}_t + e_{it} 
\end{equation}

\subsubsection{Choosing the factors}

The factor model apparatus enables dramatic reductions in dimensionality in multivariate systems, but is silent on the nature of the factors. There are broadly three approaches to this question. Firstly, the analyst can specify factors perhaps as some exogenous time series, and estimate the loadings using some variant of equation (\ref{factor-fit}). In a stock market context, a common approach is to use stock market indices, or macroeconomic time series such as interest rates or inflation as factors.

Secondly, the analyst might choose to specify the loadings or exposure matrix $\boldsymbol{\Delta}$ using pre-defined characteristics of each series. Again in a stock market context, the loadings matrix might be specified using sets of dummy variables for exposure to countries or industries, measured as an accounting ratio or perhaps some other fundamental characteristic. Given the loading matrix, the factors are estimated by cross sectional regression, period by period.

The third more `statistical' approach is to extract both the factors and the loadings from the data using a statistical model, along the lines of Principal Components or Factor Analysis (PCA/FA).

In a financial context, research has shown that the second approach, using pre-defined security characteristics can do a better job in explaining variance than a purely statistical approach \citep{Connor2019-kf} perhaps because it brings prior information regarding industry exposure or other fundamental information to bear on the problem.

Within a HF setting, it seems very natural to utilise the top few layers of the hierarchy to define the factors, as the set of aggregations are presumably chosen to reflect some sensible taxonomy of the collection of series.  Rather than define loadings as deterministic dummy variables however, we adopt a more flexible hierarchical factor model approach \cite[][and references therein]{Jorg2016-jj,Bai2015-ne} which borrows some of the PCA/latent factor apparatus. In such models, the loadings and latent factors are estimated from the data, but a degree of sparsity is imposed on the loadings matrix so that each series is exposed to a reduced set of relevant factors. Most such models require additional constraints to be imposed on the covariance structure of the factors, and we choose to allow factors within each hierarchical level to co-vary, but to be orthogonal to factors corresponding to other levels.

In a HF context, the factor model enables a particularly convenient expression for the variance of the entire collection. Using the standard HF summing matrix, the (singular) variance/covariance matrix of the entire collection can be written as:

\begin{equation}\label{var_r}
     \mathbf{S}\boldsymbol{\Delta} \boldsymbol{\Sigma}_r\boldsymbol{\Delta'}\mathbf{S'} + \mathbf{S}\mathbf{D}_r\mathbf{S'} 
\end{equation}

Stacking the $m$ univariate calibration coefficients $\rho_i$ in to a vector $\boldsymbol{\rho}$ and using (\ref{corr_g}), we can write the covariance matrix of the forecasts as:

\begin{equation}\label{var_g'}
     Var(\mathbf{g}) = diag(\boldsymbol{\rho^2})(\mathbf{S}\boldsymbol{\Delta} \boldsymbol{\Sigma}_r\boldsymbol{\Delta'}\mathbf{S'} + \mathbf{S}\mathbf{D}_r\mathbf{S'})diag(\boldsymbol{\rho^2})
\end{equation}

\subsection{Updating in theory}

Given $\underline{\mathbf{b}}$ from the exchangeable time series model, we can in theory now update the atomic level series given the base forecasts for the entire collection in one step, and aggregate via the $\mathbf{S}$ matrix to give the posterior reconciled forecasts $\bar{\mathbf{y}}$:

\begin{align}
    \mathbf{\bar{y}} &\sim N(\mathbf{S} (\underline{\mathbf{b}} + \mathbf{d}), 
    \mathbf{S} \boldsymbol{\Sigma}_d \mathbf{S'})\\
    \mathbf{d} = E(\mathbf{r}_n|\mathbf{g}) &= Cov(\mathbf{r}_n,\mathbf{g})Var^{-1}(\mathbf{g})\mathbf{g}\\
    \boldsymbol{\Sigma_d} = Var(\mathbf{r}_n|\mathbf{g}) &= Var(\mathbf{r}_n) + Cov(\mathbf{r}_n,\mathbf{g})Var^{-1}(\mathbf{g})Cov(\mathbf{g},\mathbf{r}_n)
\end{align}

\noindent where we denote by $\mathbf{r_n}$ the deviations from the prior of the $n$ atomic series. In practice, this approach is unattractive for two reasons. Computationally we need to invert the large, singular variance/covariance matrix of $\mathbf{g}$. Although some optimisations are available given a factor structure, this operation remains computationally expensive in large hierarchies due to the off diagonal terms in the idiosyncratic variance matrix $\mathbf{SD}_r\mathbf{S'}$.  Secondly the independent noise assumption is likely to be violated when considering forecasts (especially made using the same univariate models) at different levels of the hierarchy. To see this, consider a simple two level, three series hierarchy where the two atomic level series differ significantly in size. The total and the larger of the two atomic level series are likely to be strongly correlated and to generate correlated forecast errors if similar forecasting models are deployed to generate base forecasts.  

\subsection{Updating in practice}

For this reason we choose to work level by level, deriving a set of LBE updates to the entire collection of atomic level series based on forecasts from each layer of the hierarchy. That is for the set of base forecasts at hierarchical level $j$, the updated residuals at the atomic level $\mathbf{d_i}$ are distributed as:

\begin{align}
    \mathbf{d_j} &\sim N(\boldsymbol{\mu_{d_j}},\boldsymbol{\Sigma_{d_j}})\\
    \boldsymbol{\mu_{d_j}} = E(\mathbf{r_n}|\mathbf{g_j}) &=  Cov(\mathbf{r_n},\mathbf{g_j})Var^{-1}(\mathbf{g_j})\mathbf{g_j}\\
    \boldsymbol{\Sigma_{d_j}} = Var(\mathbf{r_n}|\mathbf{g_j}) &= Var(\mathbf{r_n}) + Cov(\mathbf{r_n},\mathbf{g_j})Var^{-1}(\mathbf{g_j})Cov(\mathbf{g_j},\mathbf{r_n})
\end{align}

In this setting the independence assumption is likely to be more realistic, and we proceed to combine these estimates in a second stage regression: 

\begin{align}
    \mathbf{\bar{y}} &\sim N(\mathbf{S} (\underline{\mathbf{b}} + \sum_{j=1}^k \omega_j \mathbf{d}_j), \boldsymbol{\Sigma}_{comb})
\end{align}

This gives considerable modelling flexibility, and as a bonus clearly identifies the hierarchical levels which contribute the most in terms of forecast accuracy. This can lead to improved understanding and control of the forecasting process. We then have the flexibility to modify/improve or drop entirely a set of forecasts for a given layer. This is in sharp contrast to existing approaches, which require strictly one forecast per series, and are entirely black box, producing no diagnostic information. We now turn to details of the Gibbs sampling algorithm used to reconcile the forecasts.

\section{The algorithm}

We group the full set of model parameters $\Psi$ in to blocks $\Psi=(\psi_{ts},\psi_{fm},\psi_{cal},\psi_{prop},\psi_{comb})$ corresponding to the time-series, factor model, calibration, propagation and combination stages of the model and discuss each set in turn.

\subsection{Time series models}

We choose a multivariate time series model with common parameters for all of the atomic series in the hierarchy. For the purposes of this paper, we use the exchangeable time series Dynamic Linear Model (DLM) \citep[][and references therein]{West1997}. Other suitable choices in certain contexts might be a simple random walk, as used in the Bayesian VAR literature, or the vector exponential smoothing framework with common smoothing parameters described in \cite{hyndman08}. Bayesian inference in multivariate time-series models requires estimation of the initial state variables (level and seasonality) for each series. This quite substantially increases the number of parameters in the overall model, and hence computational load. The DLM framework works particularly well in this setting as it leads to closed form estimates for these initial state parameters via backwards smoothing, substantially speeding up MCMC computation relative to exponential smoothing approaches. We omit full computational details of the model,  \citep[which are set out in great detail in][and references therein]{West1997, PradoRaquel2021}, but in summary if we were to focus on a notional univariate time-series $b$, we might write a state-space model as: 

\begin{align}
    b_t & = \mathbf{F}' \boldsymbol{\theta}_t + r_t\\
    \label{uni-state-evo} \boldsymbol{\theta}_t & = \mathbf{G} \boldsymbol{\theta}_{t-1} + w_t \\
    r_t & \sim N(0,1)\\
    w_t &\sim N(\mathbf{0},\mathbf{W}_t) 
\end{align}

With $\boldsymbol{\theta}_t$ the $p x 1$ state vector at time t, F the $p$-dimensional vector of regressors, $r_t$ the residual or observation error at time $t$, $\mathbf{G}$ the $p \times p$ state transition matrix and $\mathbf{w}_t$ the state evolution error at time $t$. The design of $\boldsymbol{\theta}$, $\mathbf{F}$ and $\mathbf{G}$ are analytical choices based on the features of the time series in question, and entail selecting the usual level, trend and seasonal time-series components. Make the standard DLM assumption that $r_t$ and $\mathbf{w}_t$ are independent both mutually and in time-series. If $\mathbf{W_t}$ are assumed known, the literature then describes the Kalman filter based steps to perform filtering, forecasting and smoothing for this model. \cite{West1997} describe a variance discounting strategy to deal with unknown $\mathbf{W_t}$, which we employ here, and show how the inference can be extended in this case.

Our main interest is in extension to the multivariate scenarios, so the model becomes:

\begin{align} \label{var_r}
     b_{t,j} & = \mathbf{F'}\boldsymbol{\theta}_{t,j} + r_{t,j} \\
     \boldsymbol{\theta}_{t,j} &= \mathbf{G} \boldsymbol{\theta}_{t-1,j} + w_{t,j}\\
     r_{t,j} &\sim N(0,\sigma^2_j)\\
     \mathbf{w}_{t,j} &\sim N(0,\mathbf{W}_t\sigma^2_j)
\end{align}

We note that $\mathbf{F}$, $\mathbf{G}$ and $\mathbf{W}_t$ are common to each series and that when $\sigma_{i,j}$ are the individual components of the error co-variance matrix $\boldsymbol{\Sigma}_r$, $Cov(\mathbf{w}_i,\mathbf{w}_j) = \mathbf{W}_t \sigma_{i,j}$, so that the states $\theta_{t,i}$ and $\theta_{t,j}$ evolve in similar (different) ways if series $i$ and $j$ are more (less) correlated. Note that the independence assumptions applied in the univariate model carry through to the multivariate case, and that the the model is defined conditional on $\boldsymbol{\Sigma}_r$. We write the states for the entire collection as $\boldsymbol{\Theta} = (\boldsymbol{\theta}_1,...,\boldsymbol{\theta}_{n})$

The parameters $\psi_{ts} = (\boldsymbol{\Theta})$ for $t=1...T$ and $j=1...n$ given  $\boldsymbol{\Sigma}_r$ are estimated and sampled using the Forward Filtering Backward Sampling (FFBS) algorithm described in detail in \cite{West1997} and \cite{PradoRaquel2021} and references therein. 

\subsection{Factor model}

We assume that $\boldsymbol{\Sigma}_r$ is well described by a factor model:

\begin{align}
    r_{t,j} &= b_{t,j} - \mathbf{F'}\boldsymbol{\theta}_{t,i}\\
    \boldsymbol{\Sigma}_r &= \boldsymbol{\Delta} \boldsymbol{\Sigma}_{f}\boldsymbol{\Delta'} + \mathbf{D_r} 
\end{align}

\noindent where $\boldsymbol{\Delta}$ is the $n x q$ matrix of factor loadings and $\boldsymbol{\Sigma_{f}}$ is the variance/covariance matrix of the $q$ factors describing the covariance of $\mathbf{r}_i$ for $i=1...n$, and $\mathbf{D}_r$ is the $n \times n$ diagonal matrix of the specific variances of $\mathbf{r}$. We choose factors which align with a subset (selected by the analyst) of the top levels of the hierarchy. The hierarchical factor model is then defined by allowing any given atomic series to load only on the factors which correspond to its hierarchical parents. In practice this is achieved using a similar technique to that underpinning Bayesian global/local shrinkage priors. For factors where a non-zero loading is appropriate, we set the prior on the elements of $\boldsymbol{\Delta}$ to have mean zero and scale of 1. Where no loading is appropriate we set the appropriate prior for $\delta_{ij}$ to have mean zero with scale also very close to zero (in practice 1e-20). Using this technique allows iterative cross-sectional/time-series regression to be used to extract the factors and loadings via Gibbs sampling.

We choose an identification scheme for the model which allows for the factors corresponding to a given hierarchical layer to be correlated with each other, whilst remaining orthogonal to factors corresponding to the other layers. This gives the factor covariance matrix a block diagonal structure. This identifies the factor model up to a sign rotation. We, therefore, impose an additional sign constraint which stipulates that each factor is positively correlated with the corresponding summed residual series from the hierarchy.

The Gibbs sampler based on this structure is efficient, and readily copes with hierarchies consisting of several hundred atomic series. Given a set of prior time-series estimates, the residuals of the atomic series $\mathbf{r}$ are calculated, the scale of these residuals is captured and used to produce a set of standardised residuals $\mathbf{r}^*$. We then fix the loadings matrix $\boldsymbol{\Delta}$ and estimate the factors $\mathbf{f}_t$ using $t=1..T$ cross sectional regressions using an independent Normal - Gamma Prior (see for example \cite{Koop2003}). For each time t, the regression equation is:

\begin{align}
    r^*_{i,t} = \mathbf{f}_t \boldsymbol{\Delta}' + e^{cs}_{i,t}
\end{align}

With standard priors for the factor and the error precision $h$:

\begin{align}
    \mathbf{f}_t|\boldsymbol{\Theta},\boldsymbol{\Delta} & \sim N(0,\mathbf{I}_q)\\
    e^{cs}_{it}|\boldsymbol{\Theta},\boldsymbol{\Delta} &\sim N(0,1/h^{cs}_t)\\
    h^{cs}_t|\boldsymbol{\Theta},\boldsymbol{\Delta} & \sim G(1,1)
\end{align}

We then impose the factor normalisation described above by first calculating the covariance matrix of the estimated factors. We then form a new `target' factor covariance matrix by setting off block diagonal terms in this empirical estimate to zero. The factors are then re-scaled (or `rotated') using the ratio of the Choleksy factors of the target matrix to the empirical matrix. That is, if the empirical covariance matrix of the estimated factors is $\boldsymbol{\Sigma}_{f0}$, we form a block diagonal matrix, $\boldsymbol{\Sigma}_{f1}$, where the between-level covariance terms are set to zero, and re-scale the estimated factors by post multiplying by $chol(\boldsymbol{\Sigma}_{f1})chol(\boldsymbol{\Sigma}_{f0})^{-1}$. This is simply a rotation of the factors and does not affect the likelihood function. We then impose the sign convention described above on the factors. We then estimate the loadings matrix $\boldsymbol{\Delta}$ (equation by equation, with coefficients as vectors $\boldsymbol{\delta}_i$) via $i=1..n$ time series regressions:

\begin{equation}
    r^*_{t,i} = \mathbf{f}_t \boldsymbol{\delta}_i + e^{ts}_{it}
\end{equation}

With priors:

\begin{align}
    \mathbf{\delta_{i,j}}|\boldsymbol{\Theta},\mathbf{F} & \sim N(0,\sigma^2_{\delta_{i,j}}) \\
    e^{ts}_{it}|\boldsymbol{\Theta},\boldsymbol{\Delta} &\sim N(0,1/h^{ts}_i)\\
    h^{ts}_{i}|\boldsymbol{\Theta}_t,\mathbf{F} & \sim G(1,1)
\end{align}

\noindent where $\sigma^2_{\delta_{i,j}}$ is set to $\sigma^2_{\delta_0}$ if series $i$ is in the part of the hierarchy exposed to factor $j$, and 1e-20 otherwise. The parameter set for the factor model component is then $\psi_{fm} = (\boldsymbol{\Delta}, \mathbf{h}^{ts},\mathbf{h}^{cs}, \mathbf{f})$ although we note that the factors and associated precisions could be integrated out of the algorithm in this stage.

\subsection{Calibration}

The next stage is to estimate the univariate calibration coefficients $\rho$ for each series. Given the samples of the states corresponding to the atomic series $\underline{b}_{t,j} = \mathbf{F'}\boldsymbol{\theta}_{t,j} + r_{t,j}$, the prior expectation of the entire collection at time t is simply $\mathbf{S\underline{b}_t}$, and we write the differences between the base forecasts $\mathbf{\hat{y}}_t$ and the prior as $\mathbf{g}_t = \mathbf{\hat{y}_t} - \mathbf{S\underline{b}}_t$. 

We run $m$ time-series regressions of $\mathbf{r}_i^*$ on $\mathbf{g}_i^*$, using a hierarchical structure to capture any commonality in predictive ability across the collection of series. The model, again using an independent Normal - Gamma prior is:

\begin{align}
    r^*_{i,t} & \sim N(\rho_i g^*_{i,t}, \sigma^2)\\
    \rho_i|\boldsymbol{\Theta}, \rho_0 &\sim N(\rho_0,\sigma^2_{\rho})\\
    \rho_0|\boldsymbol{\Theta} &\sim N(0,\sigma^2_{\rho0})\\
    h^{\rho}_i &= 1/\sigma^2 \sim G(1,1)
\end{align}

And the calibration parameter block is $\psi_{cal}=(\rho_0,(\rho_1,...,\rho_m),(h^\rho_{i},h^\rho_{i+1},...,h^\rho_{m}))$

\subsection{Propagation}

We then proceed level-by-level, the objective being to generate samples of the atomic series corresponding to the base forecasts at level $j$. We use the standardised residual forecasts $g^*$, and given ($\boldsymbol{\Theta}$, $\boldsymbol{\Delta}$ and $\mathbf{D}_r$) calculate:

\begin{align*}
    E(\mathbf{r}_n|\mathbf{g}^*_j) &= Cov(\mathbf{r}_n,\mathbf{g}^*_j)Var(\mathbf{g}^*_j)^{-1}\mathbf{g}^*_j\\
    Var(\mathbf{r}_n|\mathbf{g}^*_j) &= Var(\mathbf{r}_n) + Cov(\mathbf{r}_n,\mathbf{g}^*_j)Var(\mathbf{g}^*_j)^{-1}Cov(\mathbf{g}^*_j,\mathbf{r}_n)
\end{align*}

We assume for the purposes of this paper that these distributions are Gaussian, and draw a sample from the posterior distribution of the residuals of the atomic series updated with the forecasts from hierarchical level $j=1...k$ so that $$\mathbf{d_j} \sim N( E(\mathbf{r}_n|\mathbf{g}^*_j), Var(\mathbf{r}_n|\mathbf{g}^*_j))$$ and write $\psi_{cal}=(\mathbf{d}_1 ...\mathbf{d}_k)$.

\subsection{Combination}

The above steps result in a set of $k$ (coherent) forecasts for the entire hierarchy (i.e. one for each of the $k$ levels). We now combine those using a forecast combination style regression. Because the each of the set of $k$ forecasts is coherent, our procedure permits some flexibility in doing this; any weighted linear combination of forecasts will also be coherent, and these weights can be chosen based on combination regressions using all series or some subset of the hierarchy. In our experience, the optimal weightings are quite diffuse at the atomic level, and most reasonable weighting schemes result in substantial improvements in forecast accuracy for these series - indeed the default `sensible prior' of equal weights \citep{Elliott2013} works very well \citep{HollymanRoss2021Ufr} with no further estimation steps required. Securing forecast improvements at the top level of the hierarchy is more challenging however, so we choose these series to estimate the combination regressions. Because these series are fewer in number, focusing on the top level is also computationally more straightforward. An additional choice to be made is the number of `sets' of combination weights to be estimated. Using the atomic level of the hierarchy it is possible to estimate $k$ combination weights for each of the $n$ constituent series. At the opposite end of the scale, we can simply estimate $k$ weights for the entire hierarchy. The combination regression can be set up to achieve either, or some compromise position between the two. In our empirical example below we choose to estimate $k$ weights only for the entire hierarchy, and select the $q$ top level series corresponding to the factors in the factor model to estimate the weights. The choices here reflect the usual modelling trade-offs between estimation error and forecast accuracy. In all cases we take into account the error covariance matrix for the combination regressions. Where more than one hierarchical layer is used to estimate the combination weights, this matrix is singular. Whilst this matrix needs to be estimated, the factor model provides a strong off the shelf prior for this purpose.

In addition, as we have already taken into account the differing predictive ability of the various forecasts at each level, we constrain the weights for the combining regression to be (very close to) 1. We achieve this in a Gibbs sampler using the same mixed estimation `sum of coefficients' technique often utilised in the context of Bayesian VAR estimation \cite[see, for example,][and references therein]{Giannone2015-rh}. We utilise a Bayesian multivariate regression structure, with independent Normal - Wishart prior, where we set the prior to be an equally weighted combination of the forecasts from the $k$ different levels.

For the large hierarchy, the modelling approach is to write for each series $i=1...q$ which is being used to estimate combination weights:

\begin{equation}
    r_{it} = \mathbf{c}_{t} \boldsymbol{\omega} + e_{it}
\end{equation}

\noindent where $\boldsymbol{\omega}$ is a $k$ vector of weights/ regression coefficients, $r_{it}$ is the residual from the prior time series model at time $t$ and $\mathbf{c}_t$ is the $k$ vector of calibrated, propagated and re-aggregated base forecast residuals for series $i$ at time $t$. This can be written as a conventional SUR model with $q \times k$ regression coefficients \citep{Koop2003} modified so that $\boldsymbol{\omega}$ is constrained to be the same for each equation. We make the standard Independent Normal-Wishart distributional assumptions:

\begin{align}
    p(\boldsymbol{\omega}, H) \propto  p(\boldsymbol{\omega})p(\mathbf{H})\\
    \boldsymbol{\omega} \sim N(\underline{\boldsymbol{\omega}},\underline{\boldsymbol{\Sigma_\omega}})\\
    \mathbf{H} \sim Wishart(\underline{\nu},\underline{\mathbf{H}})
\end{align}

and add dummy observations and corresponding precisions which strongly imply $\sum_{i=1}^k \omega_i =1$ and then (omitting for clarity the dependencies on $\boldsymbol{\Theta}$ etc) sample:

\begin{align}
    \boldsymbol{\omega}|\mathbf{H} \sim N(\boldsymbol{\mu_\omega},\bar{\Sigma}_\omega)\\
    \mathbf{H}|\boldsymbol{\omega} \sim Wishart(\bar{\nu},\mathbf{\bar{H}})
\end{align}

Where posterior parameters are standard Bayesian SUR results \citep{Koop2003}.

\subsection{Prior and hyper-prior choices}

As discussed above, for time series state innovations, we employ the variance discounting strategy of \cite{West1997}. We set the discount rate for the multivariate prior model to be close to 1 (a value of 1 would imply no time variation in model parameters) so that the prior model adopts only slowly to new information. We therefore leave the base forecasts to provide rapid adaptation to changes in time series dynamics. In practice we use a value of .995. The model is robust to small changes in this parameter.

Turning to the factor model, we set the prior standard deviation on factor loadings to 1/$k_q$ where $k_q$ is the number of levels selected to be represented by factors (considering the factors and the series both standardised to have standard deviation of 1). This provides proportionally greater shrinkage when more factors are incorporated into the model, and still allows any given series to take a total unit exposure to all relevant factors.

For the calibration regressions, we set the prior mean for $\rho_0$, the system- wide calibration coefficient to zero, expressing a preference out of sample for the simpler (in terms of numbers of fitted parameters) multivariate prior. We set the prior standard deviation to .5 to allow reasonable deviation from zero where merited by the data. We then shrink individual series calibration coefficients quite strongly towards $\rho_0$, setting the prior standard deviation on $\rho_i$ to 0.1.

For the combination regressions we use an equally weighted prior of 1/$k$, with a s.d. of 2/$k$ which applies more shrinkage (to the equally weighted prior) when the number of sets of forecasts increases.  

\subsection{Gibbs sampler}

After initialisation, the sampler cycles through updates, proceeding from sample $s$ to sample $s+1$. Firstly we produce $\psi^s_{ts}|\psi^{s-1}_{fm}$, then $\psi^s_{fm}|\psi^{s}_{ts}$, before updating ($\psi^s_{calib}$,$\psi^s_{prop}$, $\psi^s_{calib}$) using values from iteration $s-1$ where necessary. In empirical work we find that by initialising the Gibbs sampler with sensible initial values for the exchangeable time series model (for which conjugate results are available), rapid convergence is achieved. We run 1000 warm-up iterations and then 2000 samples, of which we discard every other, resulting 1000 useable samples. We monitor the convergence of the Markov Chains using the Gelman-Rubin R\emph{hat} metric \citep[][and references therein]{GelmanAndrew2013Bda}, which is very close to one for all parameters and all simulations. In practice, we code our sampler in the JAX Python library \citep{jax2018github}, which provides efficient code acceleration on both CPU and GPU architectures, and sample using the distributions functions from the MCMC package Numpyro \citep[][]{phan2019composable}. We run each sample on a GPU, the calculations taking around 3 minutes per month for the relatively large Australian Domestic Tourism data set. Timings on CPU architecture are somewhat slower, but not dramatically so.

\section{Forecasting Australian domestic tourism}

We apply our model to the Australian Domestic Tourism dataset originally used in \cite{Wickramasuriya2019}. We utilise the same set of base forecasts produced by these authors and by \cite{HollymanRoss2021Ufr}. These are generated by the R Forecast package using a 96 month rolling window, leaving 131 out of sample forecasting periods. We therefore run 131 forecast reconciliation calculations. We specify an exchangeable time series model with a level component and monthly seasonal factors, and define 12 factors for our factor model; 1 factor for the whole system, 4 factors corresponding to purposes of travel (Business, Holiday, Visiting Friends and Relatives, Other) and seven other factors, corresponding to Australian states. These factors are congruent with the top three layers of the hierarchy.

\subsection{Analysis}

\begin{figure}[h]
    \centering
    \includegraphics[width=\textwidth]{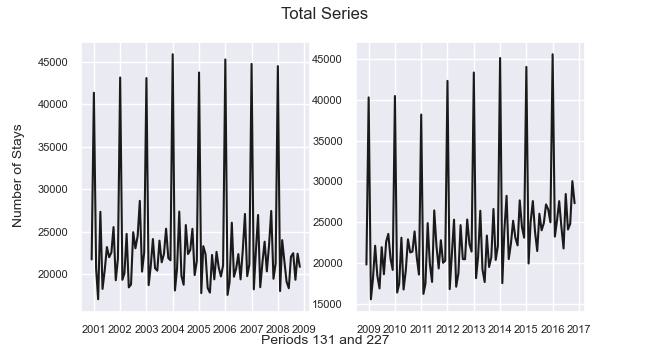}
    \caption{Total Series - time series plots}
    \label{fig:fig1}
\end{figure}

\begin{figure}[h]
    \centering
    \includegraphics[width=\textwidth]{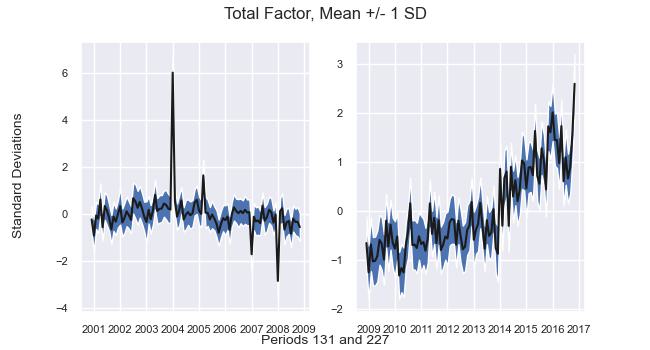}
    \caption{Total Factor - time series plots}
    \label{fig:fig2}
\end{figure}

To illustrate the difference between the aggregated time series and the corresponding latent factor, we select two (non overlapping) of the 96 month estimation windows and plot the aggregate and corresponding latent factor time-series for each. Figure \ref{fig:fig1} shows the time-series of the Total \emph{series} (measured in number of stays) for the 96 period estimation windows 131 (covering months 35 - 131; Dec 2000 to Nov 2008) and 227 (covering months 131 - 227; Dec 2008 - Nov 2016). Figure \ref{fig:fig2} plots the corresponding estimated latent `Total' factor, along with 1 standard deviation error bounds. 

Both time series plots show a relatively flat trend for the period up until 2007, a declining trend from 2007 until 2013, followed by an uptrend in the final 4 years of the period, although trends and in particular shocks are arguably clearer in the factor charts, which also have the benefit of conveying the estimation uncertainty of any particular data point. 

\begin{figure}[h]
    \centering
    \includegraphics[width=\textwidth]{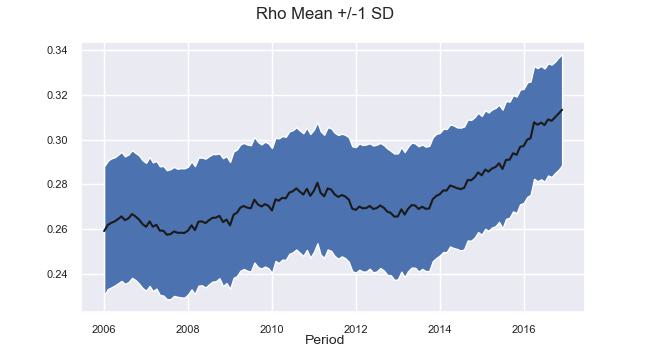}
    \caption{System-wide calibration coefficient - $\rho_0$}
    \label{fig:fig3}
\end{figure}

Figure \ref{fig:fig3} shows the system wide calibration coefficient, which we refer to as $\rho_0$. This is the target value towards which all the individual $\rho_i$ values are shrunk. We note the gentle up-trends in $\rho$ during periods where trends are discernible in the time series charts above, namely the period after the financial crisis in 2007/2008 and the recovery period commencing in 2013. This corresponds to intuition in that our prior model is specified to include a local level but not a trend component, so where there are trends in the data and these are picked up in the base forecasts, the calibration regression will tend to allocate higher weightings (higher $\rho$) to the base forecasts. Note that in our model $\rho$ is estimated using a static regression over 96 periods; it would be a relatively simple matter to use a DLM regression model to make this parameter time varying and potentially pick up on changing environments more quickly.

\begin{figure}[h]
    \centering
    \includegraphics[width=\textwidth]{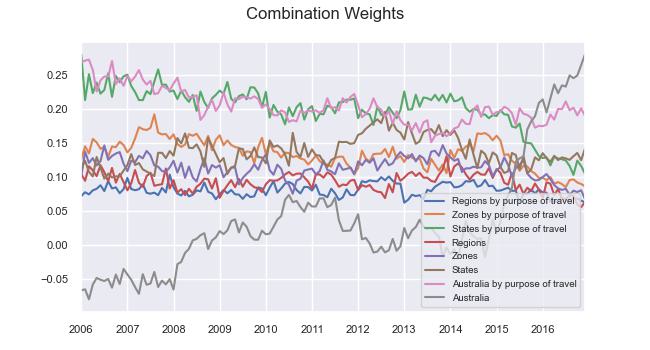}
    \caption{Combination Weights}
    \label{fig:fig4}
\end{figure}

In figure \ref{fig:fig4} we show the combination weights on the coherent forecasts derived from the set of base forecasts made at each hierarchical level. For clarity we show only the mean weight, although as with all parameters in the model, these are probabilistic. It is clear from figure \ref{fig:fig4} that forecasts by region and purpose of travel, and by state and purpose of travel tend to dominate throughout the period. Another clear feature is the increasing weight on the `Total' or `Australia' series, which starts mildly negative but ends up dominating, accounting for over 25\% of the forecast combination by the period end. Again we note that weights are is estimated using a static regression over 96 periods; DLM regression models to make this parameter time varying and potentially pick up on changing environments more quickly may further improve our forecasts. We additionally note that calibrated forecasts at the atomic level achieve consistently low weightings in the combination, suggesting that omitting to estimate base forecasts at this level and relying solely on the prior for this level may make little difference to reconciled forecast accuracy. This would have major efficiency benefits as univariate base forecasts at the atomic level are the most time consuming to produce. The weights also suggest that the `Regions' base forecasts add relatively little to forecast accuracy. We note that no other existing HF techniques facilitate this level of analysis.

\subsection{Point Forecast Accuracy}

We now turn to measures of forecast accuracy. We use an out of sample $R^2$ statistic \citep{Campbell2007-vb} to compare the point forecasting accuracy of our method to a number of alternatives, as recent research reinforces its superiority \citep{Chicco2021-ik}. The out-of-sample $R^2$ is calculated as

\begin{equation}
    R^2 = 1 - \frac{\sum_{t=1}^T (x-\hat{x})^2}{\sum_{t=1}^T (x-\bar{x})^2}
\end{equation}

With $\hat{x}$ being the forecast in question, and $\bar{x}$ the in-sample seasonal mean of the series. The out-of-sample $R^2$ statistic is not defined exclusively on $[0,1]$ in the same way as its in-sample analogue, but will give a negative value if the out-of-sample point forecast performance of the model is worse than that of the mean. We note \citep{Chicco2021-ik} that $R^2$ and MSE / RMSE have a negative monotonic relationship, so that the rankings of methodologies based on $R^2$ are the same as those based on these statistics. 

In our results we tabulate forecasts for each layer of the hierarchy from 5 different methodologies:
\begin{itemize}
    \item BASE - The (unreconciled) base forecasts for each layer
    \item OLS - Reconciled forecasts using MinT \citep{Wickramasuriya2019} with an OLS (i.e., identity matrix) covariance matrix. OLS is known to produce highly competitive reconciled forecasts for the top layers of the hierarchy. Probabilistic reconciled forecasts based on OLS were the best or  amongst the best performing in the work of \cite{PANAGIOTELIS2022}.
    \item MinTShrink - Reconciled forecasts using MinT Shrink algorithm of \citep{Wickramasuriya2019}.
    \item CCC - The combined conditional coherent approach of \cite{HollymanRoss2021Ufr}. This approach is based on simple and robust forecast combination techniques which do not depend on high dimensional covariance matrix estimation. This method tends to perform particularly well for the noisiest and hardest to forecast atomic level series
    \item HE - The `Hierarchies Everywhere' methodology described in this paper. 
\end{itemize}

The $R^2$ statistics set out in table \ref{tab1} illustrate the benefits of the hierarchical approach. There is a clear decrease in $R^2$ reading from the top left corner (most aggregate series) towards the bottom right (most disaggregate series). The HF algorithms effectively share some of the improved predictive ability evident at the top levels of the hierarchy down to the lower level series. It is also clear from the table that the HE method presented here generates more accurate point forecasts than competing approaches for most layers of the hierarchy. 

\begin{table}[h]
\resizebox{\textwidth}{!}{%
\begin{tabular}{llrrrrrrrrrrrrrr}
\toprule
& & \multicolumn{14}{c}{Forecast horizon (months)} \\
 \cmidrule(lr){3-16}
& & \multicolumn{7}{c}{All purposes} & \multicolumn{7}{c}{By purpose of travel}\\
 \cmidrule(lr){3-9}
 \cmidrule(lr){10-16}
& Approach &              1 &              2 &              3 &              6 &             12 &            1-6 &           1-12 &              1 &              2 &              3 &              6 &             12 &            1-6 &           1-12 \\
\midrule
\parbox[t]{2mm}{\multirow{5}{*}{\rotatebox[origin=c]{90}{Australia}}}
& BASE       &           53.4 &           53.5 &           50.2 &           49.7 &  \textbf{41.4} &           51.3 &           48.3 &           39.8 &           39.9 &           36.3 &           35.6 &           25.8 &           37.8 &           33.8 \\
& OLS        &           54.6 &           54.8 &           51.4 &           50.4 &           40.7 &           52.6 &           49.0 &  \textbf{40.5} &           40.2 &           37.2 &           36.1 &  \textbf{27.2} &           38.3 &           34.5 \\
& MinTShrink &           53.4 &           55.0 &           51.7 &           50.6 &           39.8 &           52.7 &           48.9 &           40.3 &  \textbf{41.2} &  \textbf{38.3} &  \textbf{36.3} &           26.4 &  \textbf{39.0} &  \textbf{34.8} \\
& CCC        &           52.3 &           54.4 &           50.9 &           49.3 &           38.1 &           51.7 &           47.5 &           37.3 &           38.1 &           35.9 &           34.4 &           25.8 &           36.4 &           32.9 \\
& HE         &  \textbf{55.0} &  \textbf{55.4} &  \textbf{52.6} &  \textbf{52.0} &           41.3 &  \textbf{53.7} &  \textbf{50.2} &           39.2 &           38.5 &           36.0 &           34.5 &           26.0 &           36.9 &           33.3 \\
\midrule
\parbox[t]{2mm}{\multirow{5}{*}{\rotatebox[origin=c]{90}{States}}}
& BASE       &           27.5 &           30.1 &           29.9 &           30.6 &           23.3 &           29.6 &           28.0 &           17.0 &           17.8 &           15.7 &           15.2 &            6.1 &           16.2 &           13.6 \\
& OLS        &           30.7 &           32.0 &           30.9 &           31.1 &           24.4 &           31.1 &           29.5 &           18.8 &           18.9 &           17.3 &           17.2 &           10.7 &           17.8 &           15.8 \\
& MinTShrink &           31.5 &           32.7 &           30.9 &           30.9 &           24.0 &           31.6 &           29.8 &           19.8 &           19.5 &           18.1 &  \textbf{18.1} &           11.8 &           18.8 &           16.7 \\
& CCC        &           31.5 &           32.7 &           30.9 &           30.3 &           23.2 &           31.4 &           29.2 &           18.8 &           18.6 &           17.8 &           17.4 &           12.7 &           18.1 &           16.5 \\
& HE         &  \textbf{32.8} &  \textbf{33.1} &  \textbf{31.5} &  \textbf{31.4} &  \textbf{25.2} &  \textbf{32.2} &  \textbf{30.4} &  \textbf{20.3} &  \textbf{19.6} &  \textbf{18.3} &           17.7 &  \textbf{13.5} &  \textbf{18.8} &  \textbf{17.1} \\
\midrule
\parbox[t]{2mm}{\multirow{5}{*}{\rotatebox[origin=c]{90}{Zones}}}
& BASE       &           12.0 &           13.0 &           12.9 &           11.6 &            6.9 &           12.4 &           11.3 &            6.5 &            6.3 &            5.8 &            4.9 &            0.8 &            5.5 &            4.1 \\
& OLS        &           14.0 &           14.6 &           14.4 &           14.5 &            9.5 &           14.4 &           13.3 &            8.5 &            8.3 &            7.7 &            7.0 &            3.0 &            7.6 &            6.3 \\
& MinTShrink &           15.6 &           16.0 &           14.9 &           15.0 &           10.0 &           15.4 &           14.1 &           10.0 &            9.8 &            9.0 &            8.5 &            4.3 &            9.2 &            7.8 \\
& CCC        &           16.5 &           17.2 &           16.5 &           16.1 &           11.6 &           16.6 &           15.3 &           10.5 &           10.4 &           10.2 &            9.8 &            6.8 &           10.2 &            9.2 \\
& HE         &  \textbf{18.5} &  \textbf{18.5} &  \textbf{17.8} &  \textbf{17.6} &  \textbf{13.6} &  \textbf{18.1} &  \textbf{17.0} &  \textbf{12.3} &  \textbf{11.8} &  \textbf{11.3} &  \textbf{10.8} &   \textbf{8.0} &  \textbf{11.5} &  \textbf{10.4} \\
\midrule
\parbox[t]{2mm}{\multirow{5}{*}{\rotatebox[origin=c]{90}{Regions}}}

& BASE       &            6.6 &            5.8 &            5.7 &            5.6 &            3.4 &            6.0 &            5.1 &            2.8 &            3.1 &            1.2 &            0.9 &           -4.0 &            1.8 &           -0.1 \\
& OLS        &            8.5 &            8.9 &            8.7 &            8.3 &            5.6 &            8.7 &            7.8 &            4.8 &            4.5 &            3.6 &            2.9 &           -0.7 &            3.7 &            2.3 \\
& MinTShrink &           10.1 &           10.8 &            9.8 &            9.2 &            5.7 &           10.0 &            8.9 &            6.1 &            5.9 &            5.1 &            4.5 &            0.9 &            5.2 &            3.9 \\
& CCC        &           12.1 &           12.6 &           12.2 &           11.8 &            8.7 &           12.2 &           11.2 &            7.9 &            7.9 &            7.8 &            7.4 &            5.3 &            7.7 &            7.0 \\
& HE         &  \textbf{13.9} &  \textbf{13.7} &  \textbf{13.3} &  \textbf{12.8} &   \textbf{9.7} &  \textbf{13.4} &  \textbf{12.5} &   \textbf{9.7} &   \textbf{9.2} &   \textbf{8.8} &   \textbf{8.2} &   \textbf{6.1} &   \textbf{8.9} &   \textbf{8.0} \\
\bottomrule
\end{tabular}}
\caption{Point forecast accuracy. The left-hand columns represent results grouped geographically. Right-hand side columns set out results for the same geographies split by purpose of travel. The data becomes more disaggregate reading downwards and to the right, with the bottom right hand side representing the most disaggregate forecasts, and overall totals in the top left corner. The numbers represent the out-of-sample $R^2$ statistic expressed as a percentage.  All base forecasts were generated using the ETS routine of the R package `forecast'.}
\label{tab1}
\end{table}

\subsection{Probabilistic Forecast Accuracy}

\cite{PANAGIOTELIS2022} examine the application of several probabilistic scoring rules to HF problems, and show that the Log Score is improper in the context of comparing unreconciled to reconciled forecasts, suggesting instead the Energy Score \citep{Gneiting2007-ya}. The Energy Score has the virtue of being easily calculated based on simulated distributions such as those produced by our Gibbs sampler, and we therefore focus on this metric to assess probabilistic accuracy.

The Energy Score for any vector $\mathbf{x}$ is calculated using 2 sets of samples (drawn with replacement) from forecast distribution of $x$, which we denote $\mathbf{\hat{y}}_1$ and $\mathbf{\hat{y}}_2$.

For a realisation $\mathbf{x}$, the energy score is:

\begin{equation}
    ES(\mathbf{x}) = E ||\mathbf{\hat{y}}_1 - \mathbf{y}||^\alpha - \frac{1}{2} E ||\mathbf{\hat{y}}_1 - \mathbf{\hat{y}}_2||^\alpha 
\end{equation}

\begin{figure}[h]
    \centering
    \includegraphics[width=\textwidth]{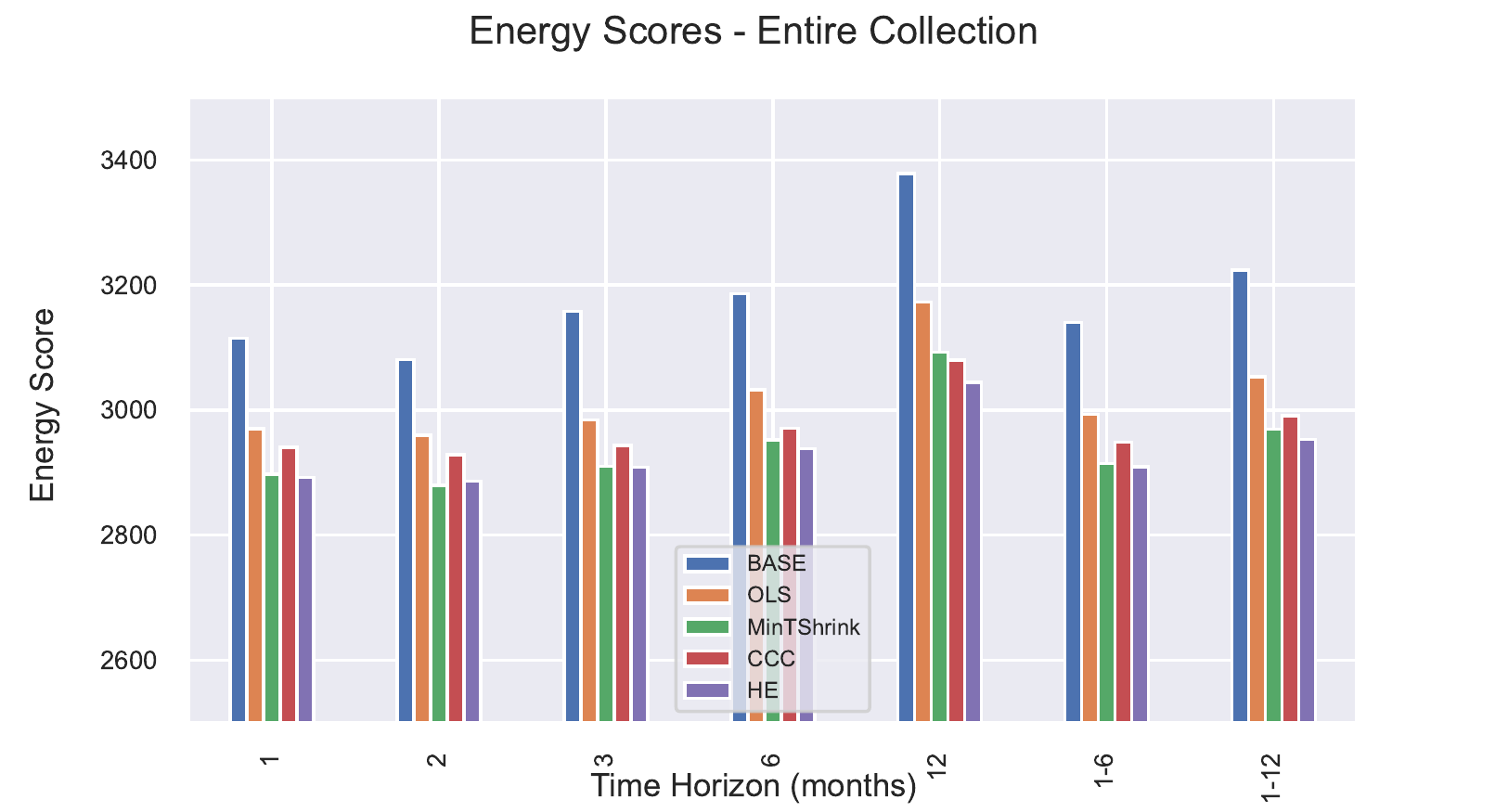}
    \caption{Energy Scores for the entire hierarchy}
    \label{fig:fig5}
\end{figure}

We calculate the energy score for the entire collection of series and then separately for each layer of the hierarchy, using $ES(\mathbf{x})$ with $\alpha=1$ as is conventional in the literature. The energy scores for the entire hierarchy are displayed graphically in figure \ref{fig:fig5}, from which we see that the HE approach outperforms the other methodologies in this exercise, with MinTShrink in second place. The OLS methodology, best performing in the work of \cite{PANAGIOTELIS2022} trails MinTShrink applied to this data set. As expected, we observe a gradually declining trend in forecast accuracy as the forecast horizon increases. 

We set out energy score results for the hierarchy layer by layer in Table 2. Although of limited relevance, as the results assess only the co-movements of series in a given level, it is of interest to note that MinTShrink performs very creditably, certainly for the top levels of the hierarchy.

\begin{table}[h]
\resizebox{\textwidth}{!}{%
\begin{tabular}{llrrrrrrrrrrrrrr}
\toprule
& & \multicolumn{14}{c}{Forecast horizon (months)} \\
 \cmidrule(lr){3-16}
& & \multicolumn{7}{c}{All purposes} & \multicolumn{7}{c}{By purpose of travel}\\
 \cmidrule(lr){3-9}
 \cmidrule(lr){10-16}
& Approach &              1 &              2 &              3 &              6 &             12 &            1-6 &           1-12 &              1 &              2 &              3 &              6 &             12 &            1-6 &           1-12 \\
\midrule
\parbox[t]{2mm}{\multirow{5}{*}{\rotatebox[origin=c]{90}{Australia}}}

& BASE       &           2,705,284 &           2,726,054 &           2,955,903 &           3,046,799 &           3,685,874 &           2,887,614 &           3,130,519 &           450,416 &           459,769 &           486,866 &           495,202 &           587,719 &           474,731 &           511,208 \\
& OLS        &           2,627,783 &           2,620,605 &           2,834,461 &           2,983,468 &           3,651,173 &           2,788,526 &           3,053,349 &           443,258 &           450,395 &           468,735 &           488,583 &           571,636 &           465,124 &           501,242 \\
& MinTShrink &  \textbf{2,576,377} &  \textbf{2,505,271} &  \textbf{2,737,313} &  \textbf{2,877,989} &  \textbf{3,606,714} &  \textbf{2,666,259} &  \textbf{2,943,967} &  \textbf{435,417} &  \textbf{433,118} &  \textbf{455,969} &  \textbf{476,734} &  \textbf{562,386} &  \textbf{450,026} &  \textbf{486,724} \\
& CCC        &           2,769,548 &           2,675,825 &           2,901,727 &           3,037,747 &           3,823,895 &           2,846,894 &           3,137,133 &           482,043 &           479,719 &           499,935 &           516,891 &           598,104 &           494,663 &           527,602 \\
& HE         &           2,645,785 &           2,638,779 &           2,833,353 &           2,906,376 &           3,659,747 &           2,761,259 &           3,016,024 &           466,185 &           475,149 &           497,351 &           513,510 &           598,325 &           489,627 &           525,079 \\

\midrule
\parbox[t]{2mm}{\multirow{5}{*}{\rotatebox[origin=c]{90}{States}}}

& BASE       &             216,513 &             209,675 &             212,746 &             212,908 &             241,561 &             213,303 &             221,003 &            44,851 &            44,679 &            46,052 &            46,384 &            52,455 &            45,720 &            47,552 \\
& OLS        &             204,432 &             202,421 &             206,169 &             209,903 &             235,450 &             206,515 &             214,553 &            43,104 &            43,282 &            43,913 &            44,595 &            48,916 &            43,947 &            45,435 \\
& MinTShrink &    \textbf{198,555} &    \textbf{195,828} &    \textbf{202,185} &    \textbf{205,980} &    \textbf{230,088} &    \textbf{200,537} &    \textbf{208,338} &   \textbf{41,920} &   \textbf{42,338} &   \textbf{42,943} &   \textbf{43,356} &   \textbf{47,107} &   \textbf{42,682} &   \textbf{44,041} \\
& CCC        &             207,279 &             203,914 &             211,360 &             215,935 &             242,799 &             209,558 &             218,874 &            44,312 &            44,425 &            45,015 &            45,607 &            49,178 &            44,892 &            46,219 \\
& HE         &             203,223 &             203,032 &             209,609 &             211,850 &             236,271 &             207,176 &             215,369 &            43,274 &            43,795 &            44,614 &            45,214 &            48,599 &            44,328 &            45,716 \\

\midrule
\parbox[t]{2mm}{\multirow{5}{*}{\rotatebox[origin=c]{90}{Zones}}}

& BASE       &              41,060 &              40,540 &              40,748 &              41,637 &              44,705 &              41,000 &              41,924 &             9,392 &             9,406 &             9,490 &             9,616 &            10,182 &             9,518 &             9,729 \\
& OLS        &              39,477 &              39,218 &              39,314 &              39,757 &              42,698 &              39,479 &              40,303 &             8,954 &             8,983 &             9,007 &             9,184 &             9,699 &             9,068 &             9,261 \\
& MinTShrink &              38,193 &     \textbf{37,863} &     \textbf{38,463} &              38,892 &              41,616 &     \textbf{38,347} &     \textbf{39,231} &    \textbf{8,700} &    \textbf{8,716} &    \textbf{8,803} &    \textbf{8,927} &    \textbf{9,430} &    \textbf{8,800} &    \textbf{8,992} \\
& CCC        &              39,132 &              38,734 &              39,162 &              39,694 &              42,522 &              39,200 &              40,127 &             9,001 &             8,998 &             9,047 &             9,147 &             9,584 &             9,057 &             9,223 \\
& HE         &     \textbf{38,165} &              38,140 &              38,540 &     \textbf{38,858} &     \textbf{41,396} &              38,461 &              39,241 &             8,817 &             8,851 &             8,922 &             9,017 &             9,440 &             8,914 &             9,079 \\

\midrule
\parbox[t]{2mm}{\multirow{5}{*}{\rotatebox[origin=c]{90}{Regions}}}

& BASE       &              12,963 &              13,012 &              13,107 &              13,201 &              13,711 &              13,068 &              13,303 &             3,115 &             3,081 &             3,158 &             3,186 &             3,378 &             3,140 &             3,224 \\
& OLS        &              12,426 &              12,336 &              12,416 &              12,586 &              13,173 &              12,442 &              12,665 &             2,970 &             2,960 &             2,985 &             3,032 &             3,173 &             2,993 &             3,053 \\
& MinTShrink &              11,951 &     \textbf{11,820} &              11,994 &              12,215 &              12,825 &              12,004 &              12,236 &             2,898 &    \textbf{2,879} &             2,910 &             2,952 &             3,092 &             2,914 &             2,970 \\
& CCC        &              12,150 &              12,025 &              12,156 &              12,278 &              12,917 &              12,161 &              12,381 &             2,940 &             2,929 &             2,943 &             2,971 &             3,080 &             2,948 &             2,991 \\
& HE         &     \textbf{11,901} &              11,875 &     \textbf{11,993} &     \textbf{12,096} &     \textbf{12,713} &     \textbf{11,978} &     \textbf{12,183} &    \textbf{2,892} &             2,887 &    \textbf{2,908} &    \textbf{2,938} &    \textbf{3,044} &    \textbf{2,909} &    \textbf{2,953} \\

\bottomrule
\end{tabular}}
\caption{Energy Scores By Hierarchical Level. The left-hand columns represent results grouped geographically. Right-hand side columns set out results for the same geographies split by purpose of travel. The data becomes more disaggregate reading downwards and to the right, with the bottom right hand side representing the most disaggregate forecasts, and overall totals in the top left corner. The numbers represent the Energy score statistics as defined in the main text for the vectors of forecasts in level k.  All base forecasts were generated using the ETS routine of the R package `forecast'.}
\label{tab2}
\end{table}

\section{Conclusion}

We introduce a novel, inherently probabilistic approach to the forecast reconciliation problem. Our approach has a number of advantages over alternatives previously proposed in the literature:
\begin{itemize}
    \item Firstly, by wrapping each stage of the procedure in to one algorithm, we enable our final forecast combination to average over uncertainty in the earlier risk and calibration stages.
    \item Secondly the approach is inherently probabilistic, and produces probabilistic reconciled forecasts whether base forecasts are probabilistic or not. This means that existing forecasting procedures can be used to generate base forecasts, which can be reconciled and turned in to distributional form in one step.
    \item Thirdly, by using a factor model we directly address the major weakness in existing approaches, namely the measurement error inherent in estimation of high dimensional covariance matrices. 
    \item Lastly we address parameter proliferation in base forecasts by shrinking predictions towards those of a model which uses a single set of time series state evolution parameters. The degree of shrinkage for each series takes in to account the data and the predictive ability of the whole collection of forecasts.
\end{itemize}
    
Our approach has two key components.  Firstly we introduce to the HF literature the idea of forecast calibration, which allows us to measure and compare the forecast accuracy of base forecasts at different hierarchical layers. The measurements produced are then used directly in the HF algorithm. Secondly we introduce a `risk model' for the hierarchy, which measures co-movement in constituent time series in a parsimonious way. As well as being an important input into our algorithm, this step can separately help to focus management attention on key elements of business uncertainty - the `risk' of any `portfolio' of atomic level series is a function of the variances of individual series, but also of the correlations between them. Assuming that the hierarchy is constructed so as to provide useful management information, choosing factors which align with hierarchical layers allows the analyst to measure risk along key dimensions of interest, and at the same time may provide valuable prior information to improve variance predictions.

For practitioners incorporating a judgemental element in their forecasting process either in the form of judgemental adjustments \citep{SEIFERT201533} or judgemental model selection \citep{PETROPOULOS201834}, both of these innovations are particularly relevant. \cite{Kremer2015} show that judgemental forecasters find it particularly difficult to account for correlations between series when making forecasts. By separating the univariate calibration process from the reconciliation step, our methodology allows for judgemental forecasts to be made on an item by item basis, and then combined with quantitative measures of risk and co-movement. 

We sketch out an alternative approach to the HF problem, which can undoubtedly be improved and developed in several directions. In particular we make several design choices in specifying the factor model which lies at the heart of the aggregated variance estimates for the hierarchy, and many alternative approaches are possible. Recent work \citep{McAlinn2017-wn} extends the multivariate BPS framework to fully time varying form, whereas our calibration equations, factor model and combination equations are currently static (non time-varying) regressions. Regarding prior choices we feel relatively strongly that the prior time series model should ideally be relatively conservative, and adapt quite slowly to new information, more of a 'long run' prior, whereas base forecasts can be more reactive. The ideal combination of both is an area for future research, as is the appropriateness of the other prior choices in terms of achieving sufficient shrinkage in different applications. Our framework holds some promise in terms of application to temporal hierarchies, and especially 'cross-temporal' \citep{Kourentzes2019,Di_Fonzo2021-tf} reconciliation, where the heuristics underlying current approaches can in theory be avoided. In the meantime, our approach represents a clear and explicable methodology to generate probabilistic forecasts for hierarchies everywhere.

\clearpage
\bibliographystyle{elsarticle-harv}  
\bibliography{references}

\end{document}